\documentstyle[epsfig]{article}

 \makeatletter
 \def\section{\@startsection {section}{1}{\z@}{3.5ex plus 1ex minus
    .2ex}{2.3ex plus .2ex}{\sc }}
 \def\subsection{\@startsection{subsection}{2}{\z@}{3.25ex plus 1ex minus
   .2ex}{1.5ex plus .2ex}{\small \sc }}
 \def\subsubsection{\@startsection{subsubsection}{2}{\z@}{3.25ex plus 1ex minus
   .2ex}{1.5ex plus .2ex}{\small \sc }}
  \def\appendix{\par\clearpage
  \setcounter{section}{0}
  \setcounter{subsection}{0}
  \@addtoreset{equation}{section}
  \def\@sectname{Appendix~}
  \def\theequation{\thesection.\arabic{equation}}
  \def\thesection{\Alph{section}}}
\makeatother
\newcommand{\be}{\begin{equation}}
\newcommand{\ee}{\end{equation}}
\newcommand{\bea}{\begin{eqnarray}}
\newcommand{\eea}{\end{eqnarray}}
\newcommand{\noi}{\noindent}

\newcommand{\pst}{\protect\textstyle\scriptscriptstyle}
\newcommand{\vr}{\mbox{\boldmath $r$}}
\newcommand{\vp}{\mbox{\boldmath $p$}}
\newcommand{\vl}{\mbox{\boldmath $L$}}
\newcommand{\vw}{\mbox{\boldmath $\omega$}}

\newcommand{\ren}{\frac{1}{2}\,L^TI^{-1}L}

\begin{document}
\thispagestyle{empty}
\parskip=12pt
\raggedbottom

\def\mytoday#1{{ } \ifcase\month \or
 January\or February\or March\or April\or May\or June\or
 July\or August\or September\or October\or November\or December\fi
 \space \number\year}
\noindent
\vspace*{1cm}
\begin{center}
{\LARGE The effect of angular momentum conservation in the phase transitions 
of collapsing systems.}

\vspace{1cm}

{\bf Victor Laliena}\\
\vspace{0.2 cm}
Hahn-Meitner Institut \\
(Bereich Theoretische Physik) \\
Glienickerstr. 100, D-14109 Berlin, Germany. 

\vspace{0.5cm}

\mytoday \\ \vspace*{0.5cm}

\nopagebreak[4]

\begin{abstract}
The effect of angular momentum conservation in microcanonical thermodynamics 
is considered. This is relevant in gravitating systems, where angular momentum
is conserved and the collapsing nature of the forces makes the 
microcanonical ensemble the proper statistical description of the physical
processes. The microcanonical distribution function with non-vanishing angular
momentum is obtained as a function of the coordinates of the particles.
As an example, a simple model of gravitating particles, introduced by 
Thirring long ago, is worked out. The phase diagram contains three phases: for
low values of the angular momentum $L$ the system behaves as the original model, 
showing a complete collapse at low energies and an entropy with a convex intruder.
For intermediate values of $L$ the collapse at low energies is not complete
and the entropy still has a convex intruder. For large $L$ there is neither 
collapse nor anomalies in the thermodynamical quantities. A short discussion
of the extension of these results to more realistic situations is exposed.
\end{abstract}

\end{center}
\eject

\section{\bf Introduction.}

Conventional thermodynamics applies to systems whose forces saturate
and therefore macroscopic parts have negligible interactions.
If this is the case,
the macroscopically conserved quantities are extensive, the thermodynamic
potentials are homogeneous functions of degree one of these quantities
and large systems can be studied in the thermodynamical limit. These
properties hold if the so called stability condition is verified. For
a system of $N$ classical particles interacting through a two body
potential, $\phi(r)$, the stability condition states that there must
exist a positive constant $E_0$ such that, for any configuration 
$\{\vr_1,\ldots,\vr_N\}$, the following inequality is obeyed \cite{linden}:

\be 
\Phi(\vr_1,\ldots,\vr_N)\;=\;\frac{1}{2}\sum_{i\neq j}\phi(|\vr_i-\vr_j|)\;
\geq -N E_0\; . \label{stc}
\ee

\noi
These cases constitute very specific situations. For example, (\ref{stc}) holds
for potentials repulsive enough at short distances, like the Lennard-Jones
potential, but  it is clearly violated for gravitating systems. When the
stability condition does not hold, the system undergoes a phase transition
from a high energy gas phase to a collapsing phase at low energies. This low
energy regime is not a proper thermodynamical phase, since it is not 
homogeneous, but, nevertheless, we shall use the name collapsing phase throughout
this paper.

Microcanonical thermodynamics has been recognized as the statistical description of 
systems which suffer fragmentation and clustering \cite{gross}, since there seems to be
problems to describe spatial inhomogeneities within the canonical formalism. These 
phenomena, fragmentation and clustering, appear in many different branches of physics,
from nuclear physics \cite{gross2} and atomic clusters \cite{mt} to 
astrophysics \cite{astro}. The task of microcanonical thermodynamics is to compute the 
entropy of a given system as a function of the macroscopically conserved quantities.
It is intuitively clear that the entropy of a system which undergoes a phase transition
associated to spatial inhomogeneities like clustering or collapsing will depend crucially
on the value of its total angular momentum, if conserved, since the spatial distribution
will do. (See ref. \cite{plasma} for a similar discussion in a different context).
Angular momentum is of particular importance in astrophysics: the formation of a single 
star or of a binary system, the birth of a solar system around a star or the
merger of galaxies in clusters is, to a large extent, determined by the value of the
angular momentum.

It is well know that, as a consequence of the virial theorem, gravitating systems
have negative specific heat \cite{pad,lb}. 
In fact, it seems that the common feature of all these collapsing systems is the 
appearance of an interval of energies where the microcanonical specific heat is 
negative \cite{th,collaps}.
Since the canonical specific heat is always positive, the microcanonical ensemble 
gives the proper statistical description of these systems. 

A system of classical gravitating particles interacting via the newtonian potential has 
an infinite entropy, due to both short and long distance singularities \cite{pad}. 
It is clear that the short
distance singularity of this potential is not physical, since new physics (like quantum
mechanics) appears at small scales. Hence, there should exist a natural short distance 
cut-off.
The long distance singularity has a different nature. In principle, one would say that
gravity should be able to bind the system. A little caveat, however, easily convinces 
oneself that, at least from the statistical point of view, the system prefers to 
evaporate than to remain bound. One needs a box (the long distance cut-off) to 
keep the system confined,
but no such a box appears in nature. We have to consider it as an artifact making
the statistical description of the system possible, which will be sensible if the 
evaporation rate is small. The box breaks translational invariance, and therefore the
momentum is not conserved. Since we are interested in keeping angular momentum conserved,
we must deal with spherical boxes, in order to maintain the rotational symmetry exact.
  
The paper is organized as follows: in section 2 we compute the 
microcanonical distribution with conserved angular momentum, by integrating out the
momenta in the phase space. The formula obtained gives a microcanonical weight
suitable for Monte Carlo simulations, and allows also for analytical approaches
relying in mean field methods. In sec. 3 we briefly derive the mean field
equations for systems which conserve angular momentum. Sec. 4 is devoted to
the discussion of a simple model, introduced by Thirring \cite{th},  which mimics the 
main features of gravitating systems with great success. The model is extended to
take into account the conservation of angular momentum. The article ends with
a summary of the conclusions, in sec 5.

\section{\bf The microcanonical distribution.}

Consider a system of $N$ classical particles whose interactions are given by a general
potential energy, $\Phi$, depending only on the position of the particles. The hamiltonian 
reads:

\be
{\cal H}\;=\;\sum_{i=1}^{N} \frac{\vp_i^2}{2 m_i}\;+\;\Phi (\vr_1,\ldots,\vr_N)\; .
\ee

If the system is isolated and the potential energy is translationally and rotationally 
invariant, the energy, momentum and angular momentum will be conserved, and, as a 
consistency condition, the center of mass will move with constant velocity. Without any
loss of generality, we can take the total momentum zero and then the center of mass
is fixed, for instance, at the origin of coordinates. Assuming some kind of ergodicity, the
microcanonical distribution will give the statistical description of the system.
The entropy $S$ is given by the Boltzmann formula:

\be
S(E,L,N)\;=\;\log \, W(E,L,N)\; ,
\ee

\noi
where $E$ is the total energy and $L$ the modulo of the angular momentum, since, by
symmetry, the entropy must be independent of the direction of the vector $\vl$;
$W$ is the volume of the phase space shell defined by the orbits with
given $E$ and $\vl$.
Notice that we took the Boltzmann constant equal to unity, and the entropy dimensionless.
Hence, the temperature is measured in units of energy.

To avoid too cumbersome expressions, let us consider the case in which only the
angular momentum is conserved. If also the linear momentum is conserved, the derivation 
of the microcanonical distribution is similar and the result will be reported at the end
of the section.
The volume of the relevant cell of the phase space can be computed starting from its
definition:

\be
W(E,L,N)\;=\;
\frac{1}{N!}\;\int\,\left(\prod_{i=1}^N\frac{d^3r_i d^3p_i}{2\pi \hbar}\right)\:
\delta(E-{\cal H})\:\delta^{(3)}(\vl - \sum_i \vr_i\wedge \vp_i)\; . \label{fundis}
\ee

\noi
Integrating out the $\vp$ variables, we will obtain a non-singular microcanonical
probability distribution depending only on the spatial configuration 
$\{\vr_1,\ldots,\vr_N\}$. In this way, we shall get an expression suitable for
microcanonical Monte Carlo simulations.

To perform easily the $\vp$ integration in (\ref{fundis}), let us define

\be 
\Pi(\bar{E},L,N,\{\vr\})\;=\;\int\,\prod_{i=1}^N d^3p_i\;
\delta(\bar{E}-\sum_i\frac{p_i^2}{2m_i})\;
\delta^{(3)}(\vl - \sum_i \vr_i\wedge \vp_i)\; , \label{integ}
\ee

\noi
where we used the notation $\bar{E}=E-\Phi(\vr_1,\ldots,\vr_N)$. It is convenient to take
the Laplace transform of $\Pi$ respect to $\bar{E}$,

\be
\tilde\Pi(s,L,N,\{\vr\})\;=\;\int_0^\infty\:d\bar{E}\:e^{-s\bar{E}}\:
\Pi(\bar{E},L,N,\{\vr\})\; ,
\ee

\noi
with  ${\rm Re}\, s > 0$. Introducing the following representation for the remaining 
Dirac delta:

\be 
\delta(x)\;=\;\int_{-\infty}^\infty\:\frac{d\omega}{2\pi}\:e^{i\omega x}\; ,
\ee

\noi
we obtain

\bea
\tilde\Pi(s,L,N,\{\vr\})&=&\int\,\frac{d^3\omega}{(2\pi)^3}\,\exp(i\vw\cdot\vl)\,
\int\,\prod_{i=1}^N d^3p_i  \nonumber \\
& &\exp\left(-s\sum_i\frac{p_i^2}{2m_i}\:-\:i\,\sum_i\vw\cdot(\vr_i\wedge\vp_i)\right)
\; .
\eea

The integral in $\vp$ is now  gaussian and can be readily performed. What shall remain 
is again a gaussian integral in $\vw$; there is no difficulty in evaluating it.
After some algebra it is found:

\be 
\tilde\Pi(s,L,N,\{\vr\})\;=\;C\,(\det I)^{-1/2}\,
\frac{e^{-s\,\ren}}{s^{(3N-3)/2}}\; , \label{tpi}
\ee

\noi
where $L^TI^{-1}L\;=\;\sum_{\alpha\beta=1}^3 L_\alpha\,I^{-1}_{\alpha\beta}\,L_\beta$ , 
the matrix $I$ is the inertial tensor respect to the origin,

\be
I_{\alpha\beta}\;=\;\sum_{i=1}^N\,m_i\,\left(\,r_i^2\delta_{\alpha\beta}\:-\:
r_i^\alpha r_i^\beta\,\right)\; , \label{it}
\ee

\noi
with $\alpha,\beta = 1,2,3$ labeling the coordinates, and
$C\;=\;(2\pi)^{\frac{3N-3}{2}}\,\prod_i m_i^{3/2}$ is a constant.

The inverse Laplace transform of (\ref{tpi}) can be found in any book
of integral transform tables \cite{tables}. It is:

\be
\frac{e^{-sb}}{s^\nu}\hspace{0.5cm}\leadsto\hspace{0.5cm}
\left\{\,
\begin{array}{ll}
0 & \mbox{\rm if  $0\,<\,\bar{E}\, <b$} \\
\frac{1}{\Gamma(\nu)}\:(\bar{E}\:-\:b)^{\nu - 1} & \mbox{\rm if $\bar{E}\,>\,b$}
\end{array} \right. \; .
\ee

\noi
Therefore

\be
\Pi(\bar{E},L,N,\{\vr\})\;=\;\frac{C}{\Gamma(\frac{3N-3}{2})}\:
(\det I)^{-1/2}\,\left(\,\bar{E}\:-\:\ren\,
\right)^{\frac{3N-5}{2}} \label{distr}
\ee

\noi
if $\bar{E}>\ren$, and it vanishes otherwise.

After the $\vp$ integration the volume of the phase space is given by:

\be 
W(E,L,N)\;=\;\tilde{C}\,\int\,\left(\prod_{i=1}^N\frac{d^3r_i}{2\pi\hbar}\right)\,
\frac{1}{\sqrt{\det I}}\left(\,E\,-\,\ren\,-\,\Phi\,\right)^{\frac{3N-5}{2}},
 \label{spatint}
\ee

\noi
where $\tilde{C}=C/[N!\Gamma((3N-3)/2)]$.
The thermodynamical quantities can be computed numerically from the last integral by, for
instance, using a suitable Monte Carlo algorithm. Eq. (\ref{spatint}) can be also used 
to derive microcanonical equations for mean field approximations.
  
If the linear momentum is also conserved, the same expressions hold, changing the exponent
in eq. (\ref{spatint}) by $(3N - 8)/2$, the constant $C$ by

\be 
C\;=\;(2\pi)^{\frac{3N-6}{2}}\,(\sum_i m_i)^{-3/2}\,\prod_i m_i^{3/2}
\ee

\noi
and the argument of the $\Gamma$ function appearing in $\tilde{C}$ by $(3N-6)/2$.
In this case, the inertial tensor $I$ refers to the center of mass, which can be taken
at the origin, and then eq. (\ref{it}) holds. Notice that, if the number of particles is 
large, the momentum conservation gives negligible differencies.

If the particles have some internal spin, then the intrinsic inertial moment of each
particle must be added to the orbital inertial moment, and the exponent in (\ref{spatint})
must be properly modified to take into account the number of intrinsic degrees of freedom.
The derivation of a microcanonical weight for these more general cases is identical to that 
outlined in this section and presents no additional difficulty. As a last remark, we
stress that the results of this section, i.e. eq. (\ref{spatint}), apply to stable as well 
as to unstable systems.

\section{\bf Mean Field theory.}

If the number of particles $N$ is very large, one can expect that the force that a 
particle undergoes will be more sensitive to the mean particle distribution of the 
system 
than to the fluctuations around it. Then one can think that the thermodynamics of 
the system depends only on the mean particle density.
In this section, we shall give a derivation of the mean field equations using this
hypothesis (cf. ref. \cite{math}). For reasons which will become clear in the following,
the results of this section apply only to unstable systems.

Let us start with the expression given by eq. (\ref{spatint}).
To perform the integral in $\vr$, we divide the integration
region in $\Lambda$ cells of volume $a^3$. The integral over $d^3r_i$ becomes a sum over
all possible cells. We have then $N$ of such sums.   
For each configuration $\{\vr\}$ we can define a function $n(i)$, with 
$i=1,\ldots,\Lambda$, which counts the number of particles in the i-th cell. It is not 
difficult to see that the initial sum over the position of the particles can be reorganized 
as a sum over the number of particles in each cell, as follows:

\bea
W(E,L,N)\;=\;A\:\sum_{n(1)=0}^N\ldots\sum_{n(\Lambda)=0}^N\,\delta\,
\left(\,\sum_{i=1}^\Lambda n(i)\,-\,N\,\right)\,\frac{N!}{\prod_{i=1}^\Lambda\,n(i)!}
\:\times\nonumber \\
\frac{1}{\sqrt{\det I}}\:\left(\,E\,-\,\ren\,-\,\Phi\,\right)^{\frac{3N-5}{2}}\; . 
\label{ds}
\eea

\noi
where $A$ is a constant which contains $\tilde{C}$, the elementary phase space volume
$2\pi\hbar$ and the volume of the cell, $a^3$.  
The combinatorial factor arises since a certain amount (= this factor) of configurations
of particle positions give the same cell occupation distribution $\{n(i)\}$.
Introducing the particle density $\rho(\vr_i) = n(i)/(Na^3)$, where $\vr_i$ denotes the
position of the center of the i-th cell, we can write (\ref{ds}) as the following 
functional integral:

\bea
W(E,L,N) \;=\; \int\,[d\rho]\:\delta\,\left(\,\int d^3r\,\rho(\vr)\,-\,1\,\right)\,
\times \hspace*{2 cm} \nonumber \\
\exp\left\{N\left[-\int d^3r \rho(\vr)\left[\log\rho(\vr)-1\right]
+\frac{3}{2}\log\left(E-\ren-\Phi\right)\right]\right\} 
\label{fi}
\eea

\noi
where we have substituted the factorials by its asymptotic form since we consider
$N$ large, and ignored some irrelevant constants. We have also neglected the term
$\sqrt{\det I}$, which is of order $1/N$. Now, we have to clarify how the argument of the
logarithm scales with $N$. For a given discretization, it is clear that the potential 
energy $\Phi$ scales as $N^2$, irrespective of whether the potential is stable or
unstable. With $N$ fixed, in the continuum limit, $a\rightarrow 0$, the potential 
energy scales with $N$ if the potential is stable,
and a proper thermodynamical limit can be taken. For unstable potentials, however, the
potential energy scales with $N^2$ still in the continuum limit.
The usual thermodynamical limit does not exist for this kind of systems. However, we can 
take $N\rightarrow\infty$ and send the masses and the coupling constants in the potential 
energy to zero, keeping $Nm=M$ and $N^2\phi(\vr,\vr')$ constant. In this case, the energy 
$E$ is not scaled; the thermodynamical functions depend on the total energy, instead of on 
the energy per particle. However, as can be seen from (\ref{fi}), the entropy scales with 
$N$. This is the scaling to be considered in this paper, and applies naturally to unstable 
systems.

The functions $I$ and $\Phi$ in (\ref{fi}) depend on $\rho(\vr_i)$ through a suitable
discretization. For very large $N$ and very small $a$, they can be expressed as:

\bea
I_{\alpha\beta}\,[\rho] & = & M\:\int\,d^3r\,\rho(\vr)\,\left(\,r^2\delta_{\alpha\beta} - 
r_\alpha r_\beta\,\right) \nonumber \\
\Phi\,[\rho] & = & \int\,d^3r\,d^3r'\,\phi(\vr,\vr')\,\rho(\vr)\,\rho(\vr')\; . 
\label{iphi}
\eea

The functional integral (\ref{fi}) is defined by the previous discretization.
There are two limits to be considered in (\ref{fi}), $a\rightarrow 0$
and $N\rightarrow\infty$. This is the rigorous order of the limits. 
For unstable systems, mean field theory consists in interchanging both
limits, taking first $N\rightarrow\infty$. In this case the functional integral (\ref{fi})
is saturated by the maximum of the exponent in the integrand, all the fluctuations around 
the maximizing density being suppressed by powers of $1/N$. 
Ignoring all the irrelevant constants, we can define the entropy per particle as

\be
{\cal S}\; =\; -\,\int\,d^3r\,\rho(\vr)\,\left[\,\log\rho(\vr)\,-\,1\,\right]\:
+\frac{3}{2}\,\log\,\left(\,E\,-\,\ren\,-\,\Phi\,\right)\; , \label{mfs}
\ee

\noi
with $I$ and $\Phi$ given by (\ref{iphi}). The physical distribution $\rho$ is that which 
maximizes ${\cal S}$ under the constraint $\int\rho(\vr)=1$.

The distribution which maximizes (\ref{mfs}) is one for which  a principal 
inertial axis goes along the angular momentum direction, for, if not, one can define
a new distribution by rotating the given distribution until the principal inertial axis
with larger inertial moment coincides with the angular momentum direction. The potential 
energy $\Phi$ and the pure entropical
term $\int\rho(\log\rho-1)$ do not change, but the rotational energy has been thus lowered, 
and hence ${\cal S}$ increases. Taking the angular momentum along the $z$ axis, we can
write

\be
{\cal S}\; =\; -\,\int\,d^3r\,\rho(\vr)\,\left[\,\log\rho(\vr)\,-\,1\,\right]\:
+\frac{3}{2}\,\log\,\left(\,E\,-\,\frac{L^2}{2I_{33}}\,-\,\Phi\,\right)\; . \label{mfs2}
\ee

\noi
The maximum can be obtained by taking the functional derivative respect to $\rho$,
which gives the following integral equation:

\be
\rho(\vr)\;=\;\exp\left\{\,\frac{2}{3}\,\beta\,\left[\,\xi\,(x^2+y^2)\:-\:
\int\,d^3r'\,\phi(\vr,\vr')\,\rho(\vr')\right]\:+\:\mu\,\right\} \; , \label{inteq}
\ee

\noi
where $\mu$ is the Lagrange multiplier for the constraint $\int\rho=1$,
and we used the notation:

\be
\beta \;=\; \left(\,E\:-\:\frac{L^2}{2I_{33}}\:-\:\Phi\,\right)^{-1}
\hspace{1cm} {\rm and} \hspace{1cm} \xi \;=\; \frac{M L^2}{2I_{33}^2}\; .
\ee

\section{\bf The Thirring model with angular momentum.}

Long ago Thirring proposed a very simple model for a star \cite{th}. In spite of its
extreme simplicity, it mimics the main features of gravitating systems with
surprisingly good success. Thirring considered the model
without angular momentum. We shall see in this section that, as expected,
angular momentum drastically changes the behaviour of the system.

\subsection{\sc The model.}

The model can be described as follows: a set of $N$ particles
are confined in a spherical volume $V$. Inside this volume there is a spherical
interaction region (core) $V_0$, concentric to $V$. Particles outside the core 
(''atmosphere'') do not 
interact, and two particles inside the core have a constant attractive potential 
energy. Using the step function

\be
\Theta_{V_0}(\vr)\;=\;\left\{
\begin{array}{ll}
1 & \mbox{\rm if $\vr\in V_0$} \\
0 & \mbox{\rm if $\vr\not\in V_0$}
\end{array} \right.
\ee

\noi
the interaction energy is given by 

\be
\phi(\vr,\vr')\;=\;=\;-\frac{Gm^2}{2}\,\Theta_{V_0}(\vr)\,\Theta_{V_0}(\vr') \; ,
\ee

\noi
where $m$ is the mass of a particle and $G$ the ''gravitational'' constant. 
(We have chosen the coupling constant in analogy with a gravitating system).
We can carry out the $N\rightarrow\infty$ limit of the previous section,
with $Nm=M$ fixed. The potential energy then reads,

\be
\Phi(\vr_1,\ldots,\vr_N)\;=\;-\frac{GM^2}{2}\,\alpha^2\; , \label{tpe}
\ee

\noi
where $\alpha =\int_{V_0}\,d^3r\,\rho(\vr)$ is the fraction of particles 
inside $V_0$.

\subsection{\sc The mean field equations.}

To simplify the computations, we shall consider the model in two 
dimensions\footnote{It should not be very difficult to solve it three dimensions, but it 
is slightly more cumbersome and no qualitative difference is expected.}. 
Notice that in this case the factor 3/2 which appears in front of 
$\beta$ in eq. (\ref{inteq}) must be substituted by 1.
Let us introduce the following notation: $R$ and $R_0$ are respectively the radius 
of the total volume, $V$, and the core, $V_0$, and $\kappa=V_0/V$.
Using the dimensionless variables $\epsilon=E/(GM^2)$ and $\Omega=L^2/(2GM^3R^2)$,
the mean field equation (\ref{inteq}) becomes:

\be
\rho(\vr)\;=\;\left\{\,
\begin{array}{ll}
\exp\,\left\{\,\mu\:+\:\beta\alpha\:+\:\beta\xi r^2/R^2\,\right\} & 
\;\;\;\mbox{\rm if $r<R_0$} \\
\exp\,\left\{\,\mu\:+\:\beta\xi r^2/R^2\,\right\}\, & \;\;\;
\mbox{\rm if $r>R_0$}
\end{array} \right. \label{mdist1}
\ee

\noi
with 

\bea
\beta &=& \left(\,\epsilon\:-\:\frac{\Omega}
{\frac{1}{R^2}\,\int\,d^2r\,r^2\,\rho(\vr)}\:+\:\frac{\alpha^2}{2}\,\right)^{-1}
\nonumber \\
\xi &=& \frac{\Omega}{\left[\,\frac{1}{R^2}\,\int\,d^2r\,r^2\,\rho(\vr)\,\right]^2}
\; .
\eea

\noi 
The two self-consistency equations

\be
\alpha\;=\; \int_{V_0}\,d^2r\,\rho(\vr) \hspace{1cm} {\rm and} \hspace{1cm}
1\:-\:\alpha \;=\; \int_{V\setminus V_0}\,d^2r\,\rho(\vr)
\ee

\noi
determine $\alpha$ and $\mu$. From them, it is straightforward to derive the
following equation for $\alpha$:

\be
\log\alpha\:-\:\log(1-\alpha)\:-\:\beta\alpha\;=\;-\beta\xi(1-\kappa)\:+\:
\log\left(\frac{1-e^{-\beta\xi\kappa}}{1-e^{-\beta\xi(1-\kappa)}}\right)\; .
\label{fe}
\ee

\noi
Notice that $\beta$ is positive, by definition. It is possible to eliminate 
$\mu$ from (\ref{mdist1}), and then the particle distribution reads

\be
\rho(\vr)\;=\;\left\{\:
\begin{array}{ll}
\frac{\alpha\,F_0}{V}\,\exp\,\left(-\beta\xi\frac{r^2}{R^2}\right)
 & \;\;\; \mbox{\rm if $\vr\in V_0$} \\
\frac{(1-\alpha)\,F_1}{V}\,\exp\,\left(-\beta\xi\frac{r^2}{R^2}\right)
& \;\;\; \mbox{\rm if $\vr\not\in V_0$} 
\end{array} \right.
\ee

\noi
where

\be
F_0\;=\;\frac{\beta\xi}{e^{\beta\xi\kappa}-1} \hspace{1cm} {\rm and} \hspace{1cm}
F_1\;=\;\frac{\beta\xi}{e^{\beta\xi}-e^{\beta\xi\kappa}} \; .
\ee

Eq. (\ref{fe}) provides the complete solution of the system. Once solved for
$\alpha$, for different values of the energy $\epsilon$ and the angular 
momentum $\Omega$, we can compute the entropy ${\cal S}$ and the temperature
$T$, which, due to the fact that the mass distribution maximizes ${\cal S}$,
is:

\be
\frac{1}{T}\;=\;\frac{\partial {\cal S}[\rho,\epsilon]}{\partial \epsilon}
\;=\; \beta \; .
\ee

Notice that $\alpha$ is an order parameter for the collapsing phase transition, 
since it is the fraction of particles inside the core $V_0$. Notice also that
(\ref{fe}) can have more than one solution for some values of $\epsilon$ and
$\Omega$. When this is the case, the true solution is that which maximizes the
entropy, the others constituting metastable states.

\subsection{\sc Limiting cases.}

Let us study the solutions of (\ref{fe}) in some limiting cases.

\begin{itemize}
\item[(a)] If $\Omega\rightarrow 0$, then $\xi\rightarrow 0$ and we recover the
original Thirring model. There is a phase transition separating a low
energy collapsing phase, where $\alpha\sim1$, from a high energy gas phase,
with $\alpha\sim\kappa$.
\item[(b)] If $\Omega\rightarrow\infty$, then $\xi\rightarrow\infty$. 
In this case $\alpha\rightarrow 0$ for $\epsilon\sim 4\Omega$ and 
$\alpha\rightarrow\kappa$ for $\epsilon\gg 4\Omega$. No collapse happens.
\item[(c)] For fixed values of $\Omega$ and $\epsilon\rightarrow\infty$, we
have $\beta\rightarrow 0$ and then $\alpha\rightarrow\kappa$. In the high
energy region, the system behaves as a gas, irrespective of the  
angular momentum, and the mass distribution is homogeneous.
\item[(d)] The most interesting case is the
behaviour of the ground state for fixed values of 
$\Omega$. In the ground state the energy of the system is 
saturated by the rotational energy (which cannot be zero due to the angular
momentum) and the potential energy, so that no energy remains for thermal
motion. This is the only microscopical contribution to the thermodynamical 
state when 
$\beta\rightarrow\infty$, and corresponds to zero temperature. It is easy to 
see that the entropy goes to $-\infty$ like $-\log\beta$, hence the 
non-thermal (rotational + potential) energy must be at its minimum, for, 
if not, the system can change its mass distribution in such a way that some
amount of thermal energy $1/\beta$ appears, rising the entropy.
In the next subsection we shall see what is the structure of the
ground state for the different values of $\Omega$.
\end{itemize}

\subsection{\sc The ground state.}

As we just discussed, when $\beta\rightarrow\infty$ the non-thermal energy

\be
\epsilon_{\pst nt}\;=\;\frac{\Omega}{\frac{1}{R}\int\,d^2r\,r^2\,\rho(\vr)}
\:-\:\frac{\alpha^2}{2}
\ee

\noi
must be at its minimum. It is possible to minimize the rotational energy without
modifying the potential energy, just taking a mass distribution with fixed
$\alpha$ which maximizes the inertial moment. Obviously, this mass distribution
correspond to a fraction $\alpha$ of the particles in the outer layer of
the core, and the remaining $1-\alpha$ fraction in the  outer layer of the
system. The dimensionless inertial moment is then 
\mbox{$(1/R^2)\int d^2r\,r^2\,\rho(\vr)=1-\alpha(1-\kappa)$}. Hence, the non-thermal
energy is

\be
\epsilon_{\pst nt}\;=\;\frac{\Omega}{1\,-\,\alpha\,(1-\kappa)}
\:-\:\frac{\alpha^2}{2}\; , \label{nte}
\ee 

The minimum of the above function is at either $\alpha=0$, $\alpha=1$, or
at one solution of

\be
\frac{\Omega\,(1-\kappa)}{\left[\,1\,-\,\alpha(1-\kappa)\,\right]^2}\:-\:
\alpha\;=\;0 \; . \label{mnte}
\ee

\noi
Eq. (\ref{mnte}) is equivalent to a real polynomial equation of third degree,
and therefore has either one or three real solutions. Since the first term
in (\ref{mnte}) has a pole at $\alpha=1/(1-\kappa)>1$,
there is always a solution at $\alpha>1/(1-\kappa)$, which is non-physical
since $\alpha$ must belong to $[0,1]$.
Besides this non-physical solution, when $\Omega$ is small (\ref{mnte})
has two more solutions. For $\Omega\ll 1$ 

\be
\alpha_{\pst max} \approx \Omega\,(1-\kappa) \hspace{1cm}
{\rm and} \hspace{1cm} \alpha_{\pst min} \approx 
\frac{1}{1-\kappa}\,\left(\,1\,-\,\sqrt{\Omega\,(1-\kappa)}\,\right) \; .
\label{minmax}
\ee

\noi
From the second derivative of (\ref{mnte}) we see that there is an inflection 
point at 

\be
\alpha_{\pst ip}\;=\;\frac{1\,-\,\sqrt[3]{\Omega (1-\kappa)^2}}{1\,-\,\kappa}
\; ,
\ee

\noi
and $\alpha_{\pst max}$ correspond to a local maximum and $\alpha_{\pst min}$
to a local minimum. We always have 
$\alpha_{\pst max}\leq\alpha_{\pst ip}\leq\alpha_{\pst min}$. Taking the 
derivative of (\ref{mnte}) respect to $\Omega$ 
- with $\alpha=\alpha_{\pst min}$ -
we easily see that $\alpha_{\pst min}$ is a monotonically decreasing
function of $\Omega$. Analogously, introducing $\alpha=\alpha{\pst min}$
in (\ref{nte}), taking the derivative respect to $\Omega$ and using 
(\ref{mnte}), we see that the value of $\epsilon_{\pst nt}$ at its local 
minimum $\alpha_{\pst min}$ is a monotonically increasing function of
$\Omega$.

Let us study the absolute minimum of (\ref{nte}) for $0\leq\alpha\leq 1$.
The three candidates are $\alpha = 0,1$ or $\alpha_{\pst min}$. For
small $\Omega$, from (\ref{minmax}) we learned that $\alpha_{\pst min}>1$,
and, since $\epsilon_{\pst nt}(0)=\Omega$ and 
$\epsilon_{\pst nt}(1)= \Omega/\kappa-1/2$, the absolute minimum correspond
to $\alpha=1$. But when $\Omega$ grows $\epsilon_{\pst nt}(1)$ increases 
faster than $\epsilon_{\pst nt}(0)$, and $\alpha_{\pst min}$ will reach the
physical region, since it decreases. The following three possibilities
define three critical values of $\Omega$:

\begin{itemize}
\item[(i)]The relative minimum $\alpha_{\pst min}$ reaches the physical
region. The critical value $\Omega_1$ is given by 
$\alpha_{\pst min}(\Omega_1)=1$. From (\ref{mnte}) we obtain
$\Omega_1=\kappa^2/(1-\kappa)$.
\item[(ii)]The relative minimum $\alpha_{\pst min}$ ceases to be the
absolute minimum. This happens for $\Omega=\Omega_2$, when
$\epsilon_{\pst nt}(\alpha_{\pst min})=\epsilon_{\pst nt}(0)$. We have
the two equations
\begin{displaymath}
\alpha_{\pst min}\;=\;\frac{\Omega_2\,(1-\kappa)}
{[\,1-\alpha_{\pst min}(1-\kappa)\,]^2} \hspace{1cm}
{\rm and} \hspace{1cm} \Omega_2\;=\;
\frac{\Omega_2}{1-\alpha_{\pst min}}\:-\:\frac{\alpha_{\pst min}^2}{2} \; .
\end{displaymath}
The solution is $\Omega_2=1/[8\,(1-\kappa)^2]$ and 
$\alpha_{\pst min}=1/[2\,(1-\kappa)]$.

\item[(iii)]The value of $\epsilon_{\pst nt}$ at $\alpha=0$ surpasses
that at $\alpha=1$. This happens when $\Omega=\Omega_3$, having
$\epsilon_{\pst nt}(0)=\epsilon_{\pst nt}(1)$. Clearly
$\Omega_3=\kappa/[2(1-\kappa)]$.
\end{itemize}

The ground state behaviour depends on the ordering of these critical $\Omega$,
which in its turn depends on $\kappa$. For $\kappa<1/2$ we have
$\Omega_1<\Omega_2,\Omega_3$. Then the ground state is characterized by 
complete collapse if $\Omega<\Omega_1$, with all the particle inside
the core. For $\Omega_1<\Omega<\Omega_2$ there is an incomplete, with 
most - but not all - the particle in the core. The value of $\alpha$
varies continuously from $\alpha=1$ at $\Omega_1$ to 
$\alpha=1/[2(1-\kappa)]$ at $\Omega_2$. For $\Omega>\Omega_2$ the ground
state has an empty core ($\alpha=0$). Notice the jump of $\alpha$ at
$\Omega_2$. 

\begin{figure}[t]
\centering
\begin{minipage}[t]{6 cm}
\setlength{\unitlength}{1cm}
\mbox{\psfig{file=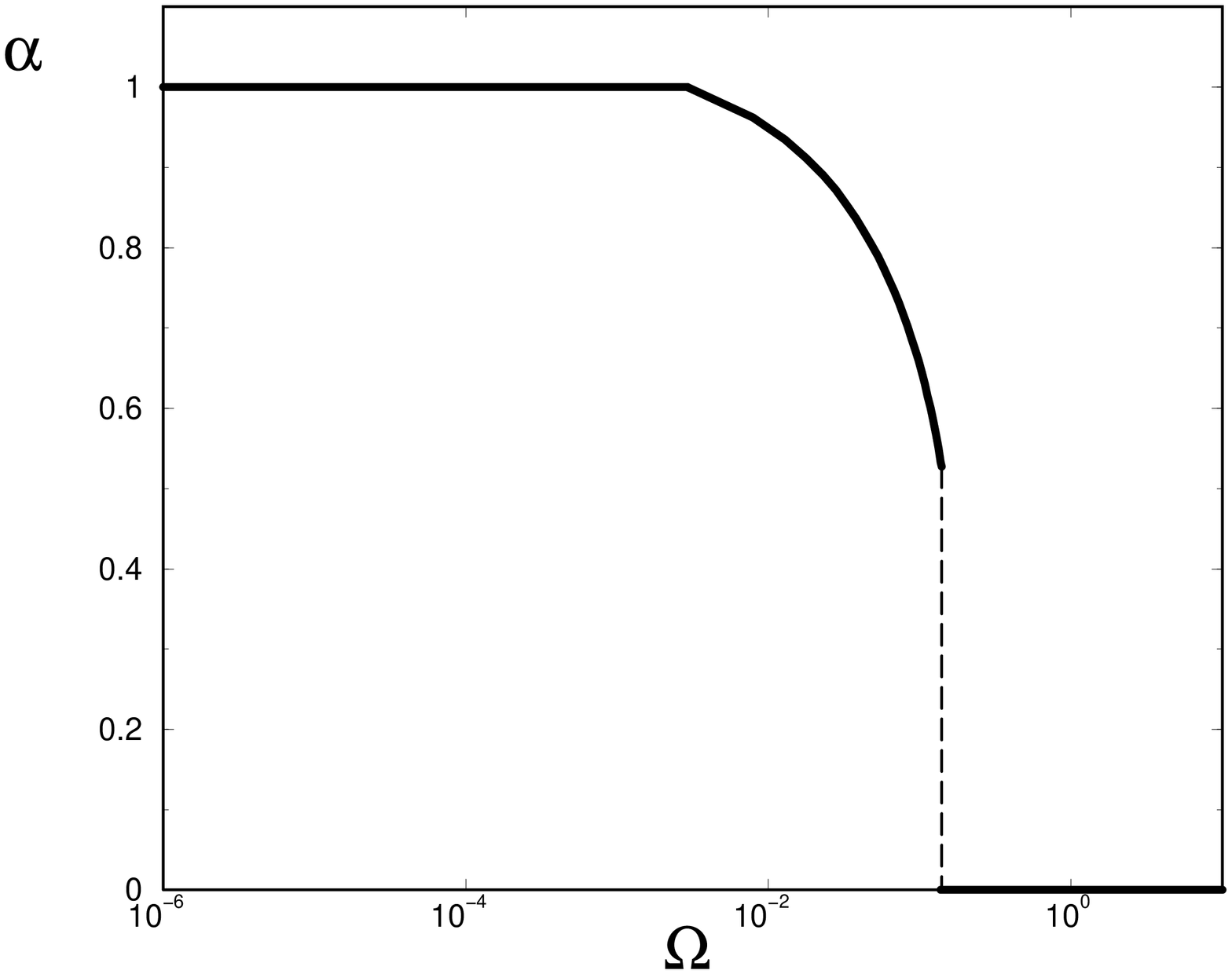,width=4 truecm, angle=-90}\put(-1,-4.3){\bf a}}
\end{minipage}
\begin{minipage}[t]{6 cm}
\setlength{\unitlength}{1cm}
\mbox{\psfig{file=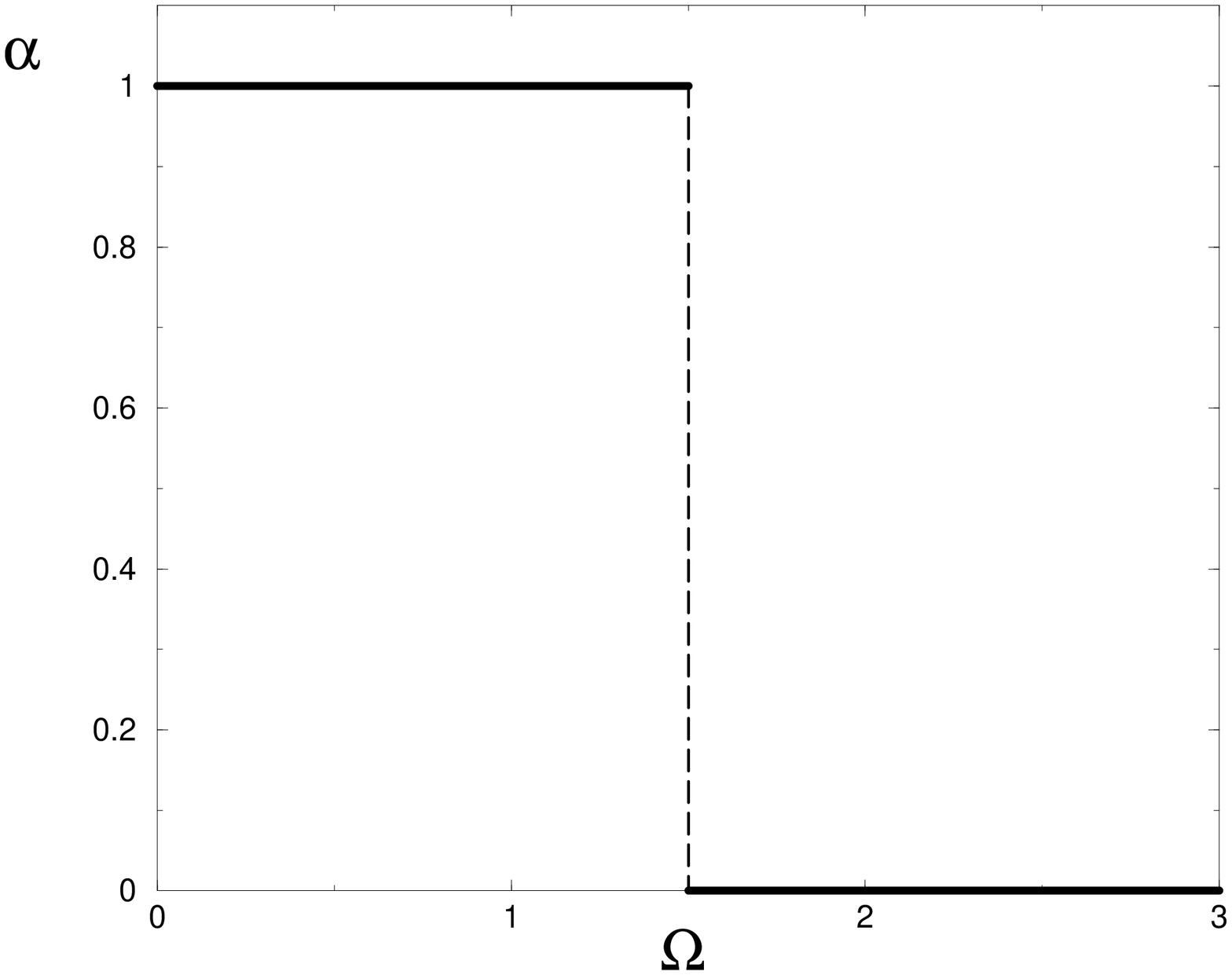,width=4 truecm, angle=-90}\put(-1,-4.3){\bf b}}
\end{minipage}
\caption{The structure of the ground state as a function of $\Omega$ for
a) $\kappa<1/2$ (observe the logarithmic scale in abscissas) and b) $\kappa>1/2$.}
\label{fig:gr}
\vspace{0.5cm}
\end{figure}

For $\kappa>1/2$ we have $\Omega_3<\Omega_1$ and $\Omega_3<\Omega_2$. 
In this case, the ground state for $\Omega<\Omega_3$ is characterized by a 
complete collapse, with all the particles inside the core. For $\Omega>\Omega_3$ 
no collapse happens ($\alpha=0$).
In fig. \ref{fig:gr} we display the behaviour of the ground state for both
$\kappa<1/2$ and  $\kappa>1/2$.

\subsection{\sc Phase diagram.}

\begin{figure}[t]
\centering
\setlength{\unitlength}{1cm}
\mbox{\psfig{file=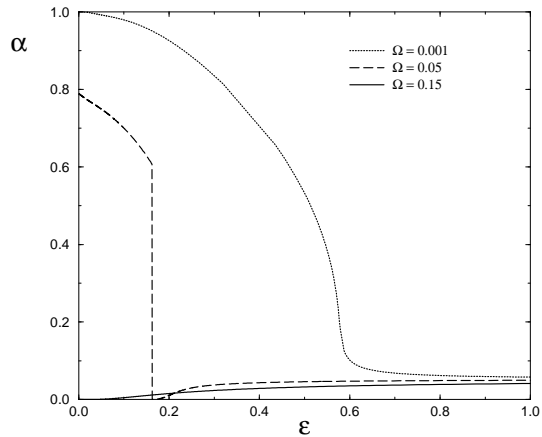,width=6 truecm, angle=-90}}
\caption{The collapsing order parameter $\alpha$ versus the energy 
for $\Omega$ values on each of the three phases. To facilitate comparisons, we 
have redefined the energy by a shift, such that the minimum energy is 
$\epsilon=0$ for all $\Omega$.}
\label{fig:alp}
\vspace{0.5cm}
\end{figure}

\begin{figure}[t]
\centering
\setlength{\unitlength}{1cm}
\mbox{\psfig{file=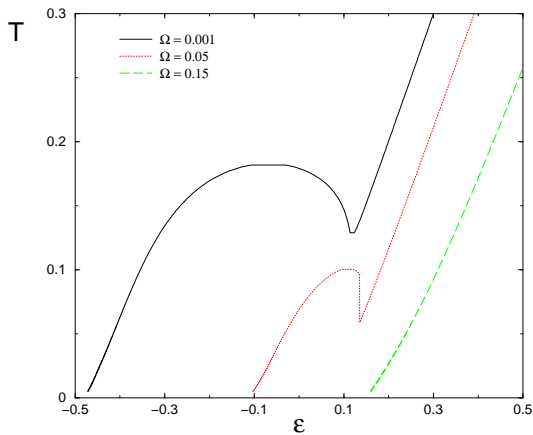,width=6 truecm, angle=-90}}
\caption{The microcanonical temperature versus the energy for three values of
$\Omega$, one on each phase of the system.}
\label{fig:tem}
\vspace{0.5cm}
\end{figure}

\begin{figure}[t]
\centering
\setlength{\unitlength}{1cm}
\mbox{\psfig{file=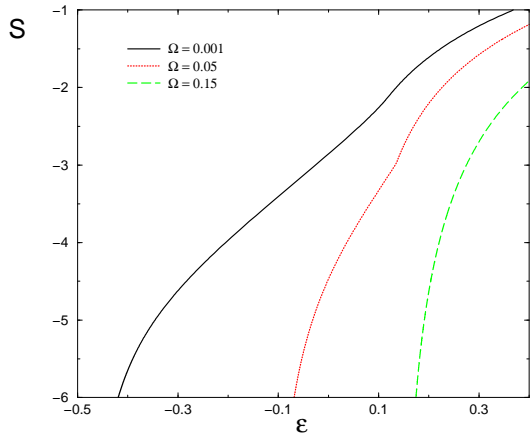,width=6 truecm, angle=-90}}
\caption{The entropy versus the energy for three values of
$\Omega$, one on each phase of the system. The convex intruder for the two smaller
$\Omega$ can be appreciated. For $\Omega=0.15$ the entropy is concave.}
\label{fig:ent}
\vspace{0.5cm}
\end{figure}

For the discussion of the model at finite temperatures we consider 
the most interesting case, $\kappa<1/2$ (atmosphere bigger than the core). 
The numerical solutions displayed in the
figures are for \mbox{$\kappa=1/(e^3-1)\approx 0.0524$}. 
 Exactly as for the ground 
state, there are three different phases, which can be distinguish by the
behaviour of the order parameter $\alpha$, which is displayed in
fig. \ref{fig:alp} as a function of the energy, for three values
of $\Omega$, representative of each phase. Figs. \ref{fig:tem} and
\ref{fig:ent} display the temperature and the entropy for the same
values of $\Omega$. It can be clearly seen the correlation between 
the collapsing transition and the anomalies in the thermodynamical
quantities. Let us discuss in detail what is happening on each phase.

\begin{itemize}
\item[(1)]
 $\Omega<\Omega_1$. There is a complete collapse at low energies, with
$\alpha\approx 1$. The gas phase at high energies is separated from the
collapsing phase by an interval of energies with negative specific heat.
Qualitatively, the model is similar to the original ($\Omega=0$) Thirring
model. The entropy shows the characteristic convex intruder in the energy 
interval where the two phases coexist. (Notice that, since the system is not
thermodynamically stable, van Hove's theorem \cite{vanh} does not apply).
\item[(2)]$\Omega_1<\Omega<\Omega_2$. At low energies the collapse is not
complete, with $\alpha<1$. For values of $\Omega$ near $\Omega_1$ the 
thermodynamical quantities are qualitatively equal to those in the low
angular momentum phase (negative specific heat).
When $\Omega$ is larger, eq. (\ref{fe}) has more than one solution.
For energies larger than a certain value, one of the new solutions becomes 
the absolute maximum of the entropy. This is the origin of the jumps in
$\alpha$ and in $T$ that can be seen in figs. \ref{fig:alp} and 
\ref{fig:tem}, for $\Omega=0.05$. Notice that the jump occurs after
a region with negative specific heat. For $\Omega$ still larger, the jump
appears before the specific heat becomes negative. However, although in the last 
cases the specific heat is positive everywhere, the entropy still has a convex 
intruder, due to the kink originated by the jump in the temperature.
\item[(3)]$\Omega>\Omega_2$. There is no collapse at low energies. The
specific heat is positive, smooth and increases monotonically with the
energy. The entropy has no convex intruder.
\end{itemize}

\begin{figure}[t]
\centering
\setlength{\unitlength}{1cm}
\mbox{\psfig{file=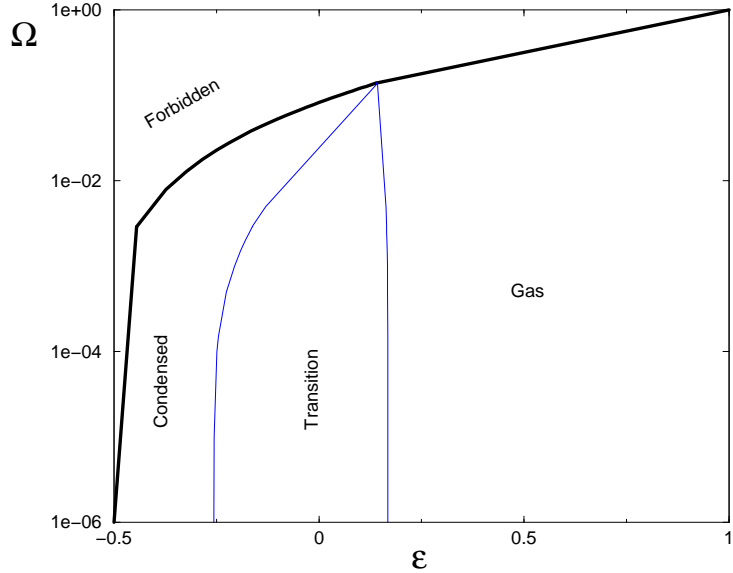,width=8 truecm, angle=-90}}
\caption{The phase diagram of the model. Notice the logarithmic scale in the 
ordinate axis. The thick line corresponds to the $T=0$ isotherm.}
\label{fig:phd}
\vspace{0.5cm}
\end{figure}

From the above discussion, it becomes manifest the correlation between the 
collapsing phase transition and the anomalies in the caloric curve ($T$ versus 
$\epsilon$) and in the entropy. 
The phase diagram in the plane $(\epsilon,\Omega$ is displayed in fig. 
\ref{fig:phd}.
Obviously, there is a forbidden region, where the system cannot be since a minimal
rotational energy is required to keep angular momentum constant.
The boundary between the forbidden and allowed regions is the $T=0$ isotherm. 
Below the forbidden region,
there are two phases\footnote{The two collapsing phases, $\Omega<\Omega_1$ and 
$\Omega_1<\Omega<\Omega_2$, are distinguished only by the behaviour of the order
parameter at low energies. The thermodynamical quantities, however, behave
in a similar way in both cases. Therefore, in the phase diagram we shall not 
distinguish the two low energy phases.}  separated by a transition region, which 
is defined through the Maxwell construction. Since both phases are
qualitatively different, they must be analytically separated. This means that
the critical point must be at zero temperature (i.e., on the curve of minimal 
energy). And, in its turn, this implies that the critical point must be at
$\Omega_c=\Omega_2$, since this is the point on the zero temperature line where
the collapse disappear. This is what actually happens, as can be seen in fig. 
\ref{fig:phd}.

For fixed $\Omega$, the difference of the energies limiting the transition
region defines the latent heat. We see that the phase transition is first
order everywhere, with non-vanishing latent heat. From the analysis
of eq. (\ref{fe}) when $\beta\rightarrow\infty$ and $\Omega\rightarrow\Omega_2$
one concludes\footnote{After a careful study of the possible solutions, their
entropies, etc., which is not reported here in detail.} that the latent heat
vanishes as $\Omega_2-\Omega$.
Therefore, at the critical point, $\Omega_2$, the latent heat disappears. 
Since the specific heat is continuous at this point, we cannot say  
that the phase transition becomes second order; indeed, 
the order parameter $\alpha$ jumps at $\Omega_2$.

\subsection{\sc Discussion.}

Now, we want to address the
question of whether the phase diagram described in the previous subsection 
shares its main properties with those of more realistic models, or whether
these are merely a consequence of the extreme simplicity of the model .
We can give at least a partial answer.
The model considered here is unrealistic mainly due to the fact that the 
particles can only collapse in a fixed region. Therefore, there is no place
here for more complicated regimes like multifragmentation. 

The collapsing phase, at low angular momentum $L$, with all - or most - the 
particles forming a cluster, is expected to be present in realistic problems.  
Of course, the size of the cluster will increase with angular 
momentum. For large energies both the potential and rotational energy
are negligible and we shall have a gas phase. There will be then a 
transition region separating this two phases, like in fig. \ref{fig:phd}.
The difference may be, however, that in realistic cases these two phases
might be connected through intermediate regimes like
multifragmentation. It is clear that for large enough angular momentum
the system will prefer to collapse into several bodies rather than in a
single one. The critical point, if any, might be located at non-zero
temperature. If this is the case, one could expect a true second order
phase transition, with divergent specific heat and large fluctuations
of some local order parameter, like the density. Needless to say, these
are questions which deserve a careful investigation.

\section{\bf Conclusions.}

Let us briefly summarize the results presented in this work. First,
we found a non-singular expression for the microcanonical distribution of  
systems with conserved, non-vanishing angular momentum. It was obtained by
integrating out the momenta in the phase space, and it is suitable for
Monte Carlo simulations, as well as for mean field computations.
We  presented a derivation of the mean field equations for systems of
classical particles interacting through a two body unstable potential, 
taking into account the conservation of angular momentum.

We discussed, as an application, the properties of the phase diagram of a 
simple model of gravitating particles with non-vanishing angular momentum. 
We found the exact solution in the mean field approach, which served to 
illustrate how the phase diagram of a physical system can be modified by 
the conservation of the angular momentum. In the case under interest, the 
angular momentum $L\sim \sqrt\Omega$ determines the existence of three phases, 
separated by two critical values $\Omega_1$ and $\Omega_2$. 
At low energies all particles condense in a small region if $\Omega<\Omega_1$
For $\Omega_1<\Omega<\Omega_2$ the collapse at low energy is incomplete
with most - but not all - particles condensed in a small region. At high
energies, the system behaves like a gas irrespective of the value of $\Omega$.
The collapsing-gas phase transition for $\Omega<\Omega_2$ is accompanied by
an anomaly in the caloric curve ($T$ versus $\epsilon$), which reflects the
fact that the entropy is not concave in an energy interval.
This interval coincides with that where the collapsing-gas transition
takes place. Through the Maxwell construction, it is possible to
define precisely this transition region and a latent heat, which shows that 
the transition can be classified as first order. The latent heat disappear 
continuously at $\Omega=\Omega_2$ and $T=0$. This reflects the fact that both 
phases, being qualitatively different, cannot be analytically connected in the 
phase diagram.
At $\Omega>\Omega_2$ there is neither collapse nor anomaly in the 
thermodynamical quantities. We argued that, with more realistic potentials, 
an intermediate regime between the collapse and gas phases - multifragmentation 
perhaps - should exist. Therefore, it might be possible that the system pass
smoothly from one phase to the other. Therefore, the critical point might be
located at non-zero temperature, showing the typical behaviour of a second 
order phase transition.

\noi {\sc Acknowledgements.}

The author is grateful to D.H.E. Gross for driving his attention to the problem
of angular momentum in microcanonial thermodynamics and in gravitating systems. 
He wants to thank O. Fliegans, 
D.H.E. Gross and Th. Klotz for stimulating discussions. He is also indebted with 
O. Fliegans for providing him some interesting references, with Th. Klotz for 
technical help and with E. Follana for pointing out some misprints in the first
version of this manuscript.

\end{document}